**Title: Are foundation models efficient for medical image segmentation?**

Authors: Danielle L. Ferreira PhD[1], Rima Arnaout MD[2]*

[1]Department of Medicine, Division of Cardiology
Bakar Computational Health Sciences Institute
University of California, San Francisco
521 Parnassus Avenue
San Francisco, CA 94143

[2]Department of Medicine, Division of Cardiology
Bakar Computational Health Sciences Institute
UCSF-UC Berkeley Joint Program in Computational Precision Health
Department of Radiology, Center for Intelligent Imaging
University of California, San Francisco
521 Parnassus Ave Box 0124
San Francisco, CA 94143
* Corresponding Author

Word Count: 1209

**Abstract**

Foundation models are experiencing a surge in popularity. The Segment Anything model (SAM) asserts an ability to segment a wide spectrum of objects but required supervised training at unprecedented scale. We compared SAM's performance (against clinical ground truth) and resources (labeling time, compute) to a modality-specific, label-free self-supervised learning (SSL) method on 25 measurements for 100 cardiac ultrasounds. SAM performed poorly and required significantly more labeling and computing resources, demonstrating worse efficiency than SSL.

With larger and more capable foundation models than ever before, many are asking whether, and how, they will accelerate AI for medicine[1].

Clinical image segmentation, a critical task across medicine, is in desperate need of greater efficiency—better performance with less manual work. For example, manually tracing cardiac chambers to generate structural and functional measurements is clinically important[2] but also exceedingly laborious and error-prone. Deep learning holds potential to automate segmentation for greater efficiency, and foundation models like SAM have now been proposed for this task.

SAM is a supervised model trained on over 1 billion manual/semi-manual annotations[3]. SAM is designed to segment new objects either automatically, or, in an interactive manner through manual prompts. Prompting SAM greatly improves segmentation performance[4] and consists of three steps (example in **Figure 1**): (i) clicking points and/or drawing bounding boxes on image regions of interest, (ii) choosing which of SAM's attempted segmentation masks corresponds best with the object(s) of interest, and (iii) choosing which segment(s) correspond with meaningful object(s), i.e. providing semantic information. SAM can be deployed on medical imaging in a one-shot approach or, after additional training (fine-tuning) on modality-specific images.

Testing SAM on medical imaging for several anatomic structures across imaging modality has already begun[4–6], including on cardiac ultrasound[5,6]. For example, Chao et. al. performed zero-shot evaluation of SAM on segmenting the left ventricle from publicly available cardiac ultrasound images and reported an acceptable average Dice score of 0.86; of note, fine-tuning did not improve SAM's performance[6]. These same research papers found that SAM displays heterogeneous performance on medical imaging, often underperforming compared to traditional supervised segmentation[4,5,7]. Critically, it must be noted that current reports evaluating SAM for medical imaging[4,5,7,8] use prompts either manually generated at the point-of-use or derived from manual annotations from already-labeled datasets for tasks (i)-(iii) mentioned above. For example, in Chao et al., manual tracings of the LV were used to create bounding box prompts specific to the LV, and then again to choose the segment best corresponding to the object of interest (by Dice score with manual label) and provide semantic information. Therefore, SAM still relies on laborious human labels for inference.

One alternative to supervised foundation models is task-specific, label-free models. For example Ferreira et. al. leveraged computer vision, clinical knowledge, and deep learning to provide self-supervised (SSL) cardiac ultrasound segmentation with no human labels[9]. Here, we sought to compare efficiency—performance and resources—of SAM vs SSL for segmentation of cardiac ultrasound.

Briefly, we used the SSL model to segment cardiac chambers from images from 100 cardiac ultrasounds (1350 images) across a range of shapes, sizes, and patient characteristics (Table S1) as previously described[9]. One-shot SAM (original SAM ViT-H model having 636M parameters, model type "vit_h", checkpoint "sam_vit_h_4b8939.pth", https://github.com/facebookresearch/segment-anything) was used to segment the same images. (As above, one-shot SAM has been reported to function similarly to fine-tuning for cardiac ultrasound[6].) Because automatic (no prompts) SAM performs unacceptably poorly[4], box prompts were used to manually target each chamber of interest—notably, prompts were independently performed by a clinician rather than derived directly from the same label used for comparison of SSL and SAM. A clinician was again used to select the best chamber mask from SAM's three predictions.

Chamber segmentations were used to calculate clinical biometrics in accordance with clinical guidelines and as previously described[2,9]. Statistical comparisons were also performed as before[9]; for statistical comparison of Bland-Altman biases and limits of agreement with different units of measure, measurement differences between model and ground truth were converted to percentages. Deidentified patient imaging was used with waived consent in compliance with the UCSF IRB. All inference was done locally; no patient data was submitted to SAM via the internet. An overview of these methods is found in **Figure 1**.

SSL- and SAM-derived cardiac measurements were each compared against measurements from the clinical report as ground truth. Determination coefficient ($r^2$), Bland-Altman bias±limits of agreement (LOA – two standard deviations) were measured (**Figure 2**). SAM had low correlation and bias±LOA compared to clinical measurements, with mean $r^2$ of 0.39±0.16 (range, 0.08-0.69), mean bias 17%±21% (range, -23% to 81%), and mean LOA of 86%±34% (range 42%-172%). SSL had a mean $r^2$ of 0.75±0.12 (range, 0.54-0.93), mean bias 1%±17% (range, -31% to 38%), and mean LOA of 47%±18% (range, 26%-95%). By all measures, SSL outperformed SAM (Mann-Whitney U p-values for $r^2$, p<1e-07; for bias, p=0.01, and for LOA, p<1e-05). Only two of 25 SAM-derived measurements had an $r^2$ of greater than 0.6, while 20 of 25 SSL-derived measurements had $r^2$>0.6.

We also measured resources used to train and perform inference for SAM and SSL (**Table 1**). While trained for a broader set of objects on a single modality (natural images), SAM required over 100 times more images to train than SSL. SAM requires 1.1 billion labels which were performed manually and semi-manually (number of human labeling hours was not available)[3], while SSL required no manual labels. Prompting time was also measured; SSL requires no prompting to provide inference, while SAM required over 4 hours of manual prompting time total for only 100 cardiac ultrasounds (2.46 minutes/ultrasound). Model size, required training GPUs, and inference time per image were all greater for SAM than for SSL.

A major reason deep learning is of interest for medical imaging is the potential to achieve greater efficiency for clinicians' most difficult and laborious tasks, such as medical image segmentation. Efficiency is in turn a function of performance attained for resources—especially manual human resources—expended.

While SAM may segment objects in natural images well, this comes at significant cost, both up-front during model training as well as at inference time via human prompting. For example, compared to the original Ferreira et. al. which evaluated 18,423 ultrasounds, such an evaluation using SAM would have taken over 755 human prompting hours. For a specific, clinically important, and difficult medical imaging task such as segmentation of the heart in cardiac ultrasound, the smaller, modality-specific SSL model performed better against clinical measurements, required fewer computational resources and no manual labeling neither during initial training nor at inference time.

While traditional supervised segmentation has often outperformed SAM to date[7], SSL provides further efficiency by maintaining performance without manual labels. Fine-tuning SAM would require even more manual and computational resources and has not outperformed one-shot SAM on this task to date[6]. Large foundation models may also be less interpretable than task-specific models, a concern that may grow as training datasets and full algorithmic details for foundation models are not always forthcoming[10]. Foundation models are an exciting avenue for deep learning; however, maintaining focus on efficiency is still key for medical tasks. Moving away from supervised models of all sizes can help further efficiency if clinical performance is maintained. Exploration of several types of deep learning will likely benefit the field and its applications for medical imaging.

**Code Availability.** Code will be made available upon publication at github.com/ArnaoutLabUCSF/CardioML

**Data Availability.** A minimum dataset is available at https://www.creatis.insa-lyon.fr/Challenge/camus/index.html

**Author Contributions.** R. A. conceived of the study. D.F. performed all analyses. Both authors wrote the manuscript.

October 24, 2023. https://www.wired.com/story/fast-forward-ai-powerful-secretive/

**Tables**

**Table 1. Human and computational resources required by SSL vs. SAM.**

|  | SSL | SAM |
|---|---|---|
| **Nb of training images** | 93,000 | 11M |
| **Nb of manual/semi-manual labels** | 0 | 1.1B |
| **Prompting time (manually bounding boxes)** | 0 | 3 hours (Nb = 1350 masks) |
| **Mask selection time** | 0 | 1.1 hour (Nb = 4050 masks) |
| **Model size** | 93 MB | 2.56 GB (SAM-h) |
| **Parameters** | 7.7M | 636M |
| **Computational resource for training** | 1 GPU | 256 GPUs |
| **Inference time per mask** | 0.1 sec | 2.7 sec |

**Figure Legends**

**Figure 1. Manual input needed for SAM vs SSL inference.** SAM inference steps (top) include (i) creating bounding boxes (green) for each chamber, (ii) selecting the best prediction (green) from among three options SAM provides per object, and (iii) assigning semantic information to the resulting segments to achieve the final segmentation result. (B) SSL predicts all segmentation masks simultaneously as previously described[9] without manual inputs.

**Figure 2. Comparison of SSL and SAM to clinical ground-truth.** (a) Correlation ($r^2$) between model-derived chamber measurements and clinical ground truth for 25 cardiac measurements. White circles, SSL model comparison against clinical ground truth; pink circles, SAM. Vertical shading indicates $r^2$ ranges for poor (purple), fair (blue), moderate (teal), good (green), and excellent (yellow) agreement. (b) Bias and limits of agreement (LOA) comparison of clinical vs. SSL- and clinical vs. SAM- derived chamber measurements. Open circle, bias. Blue, SSL. Orange, SAM. Darker color, one standard deviation. Lighter color, two standard deviations (LOA). Dashed line at bias=zero. LV, left ventricle. Endo, endocardium. Epi, epicardium. LVEDV, LV end-diastolic volume. LVESV, LV end-systolic volume. LVEF, LV ejection fraction. LA, left atrium. LAEF, LA ejection fraction. RV, right ventricle. RA, right atrium.

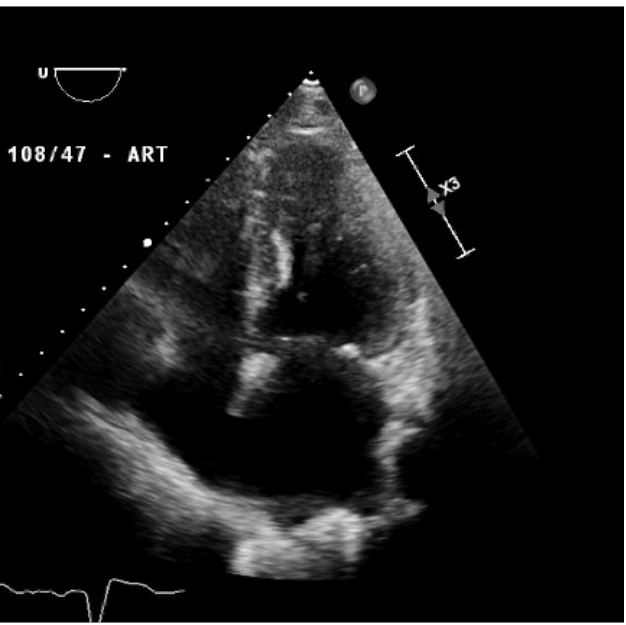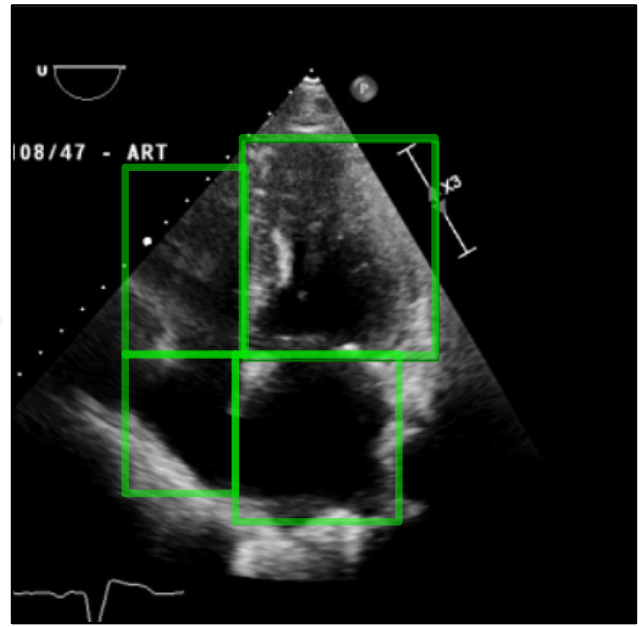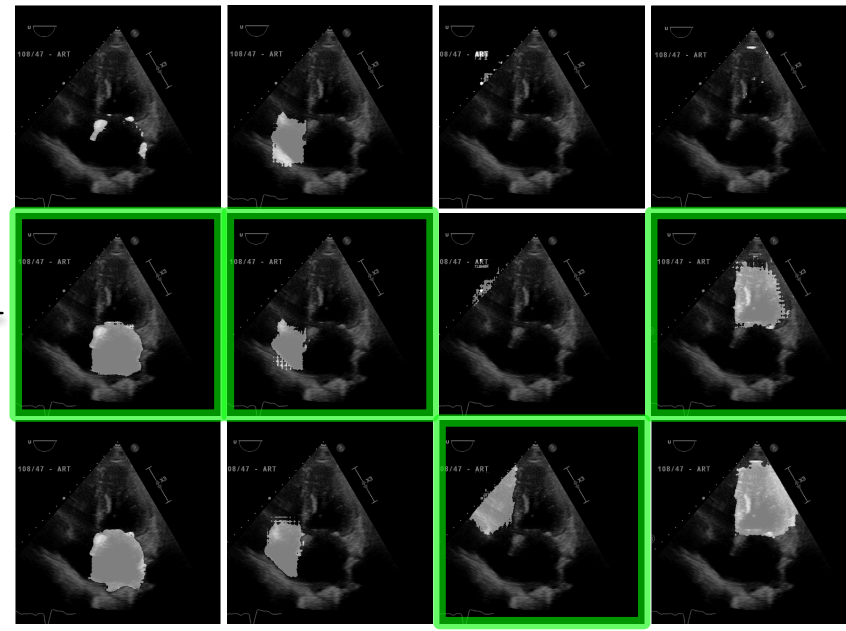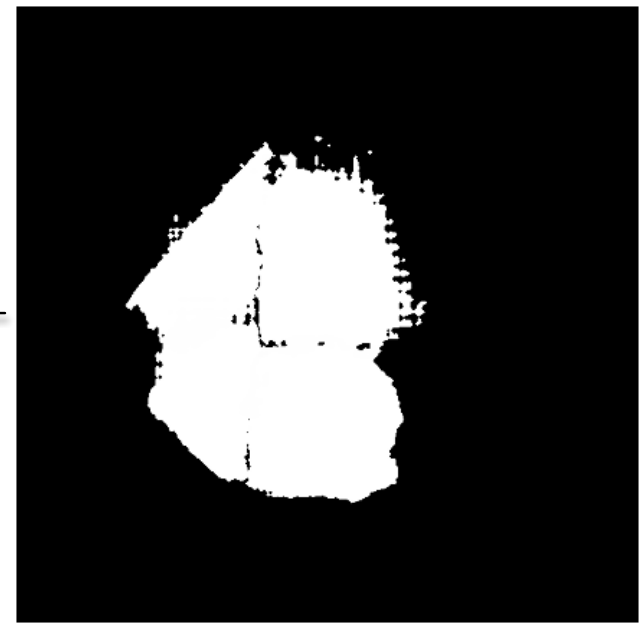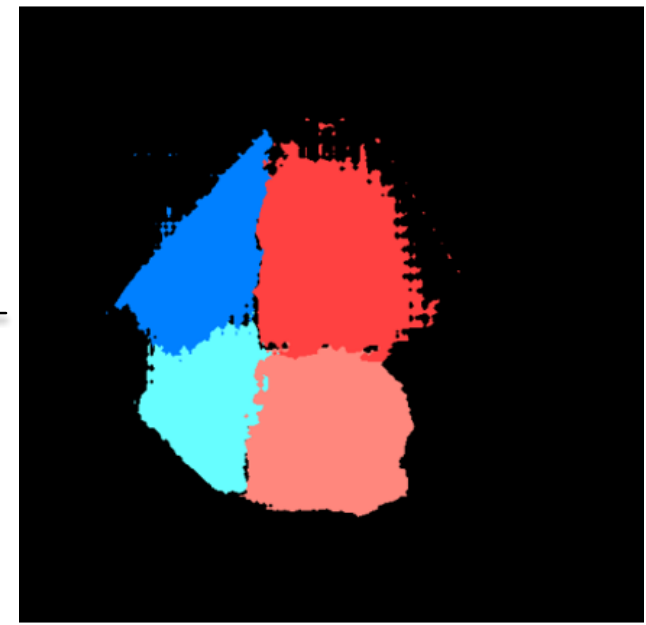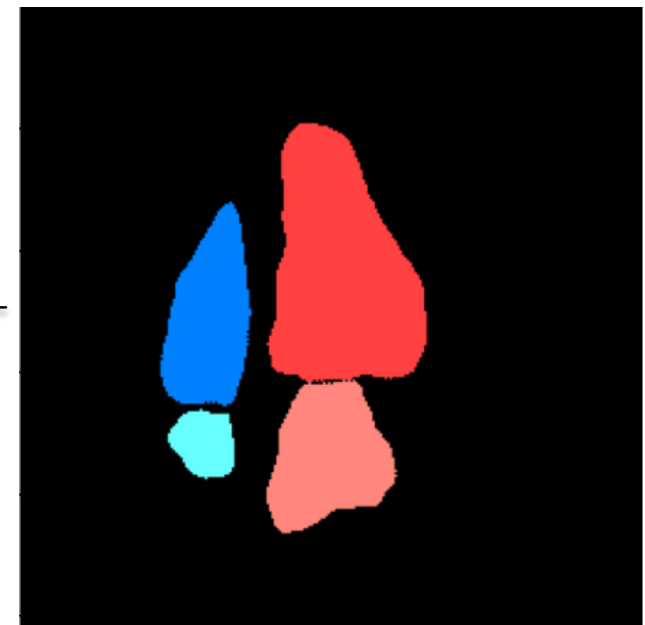

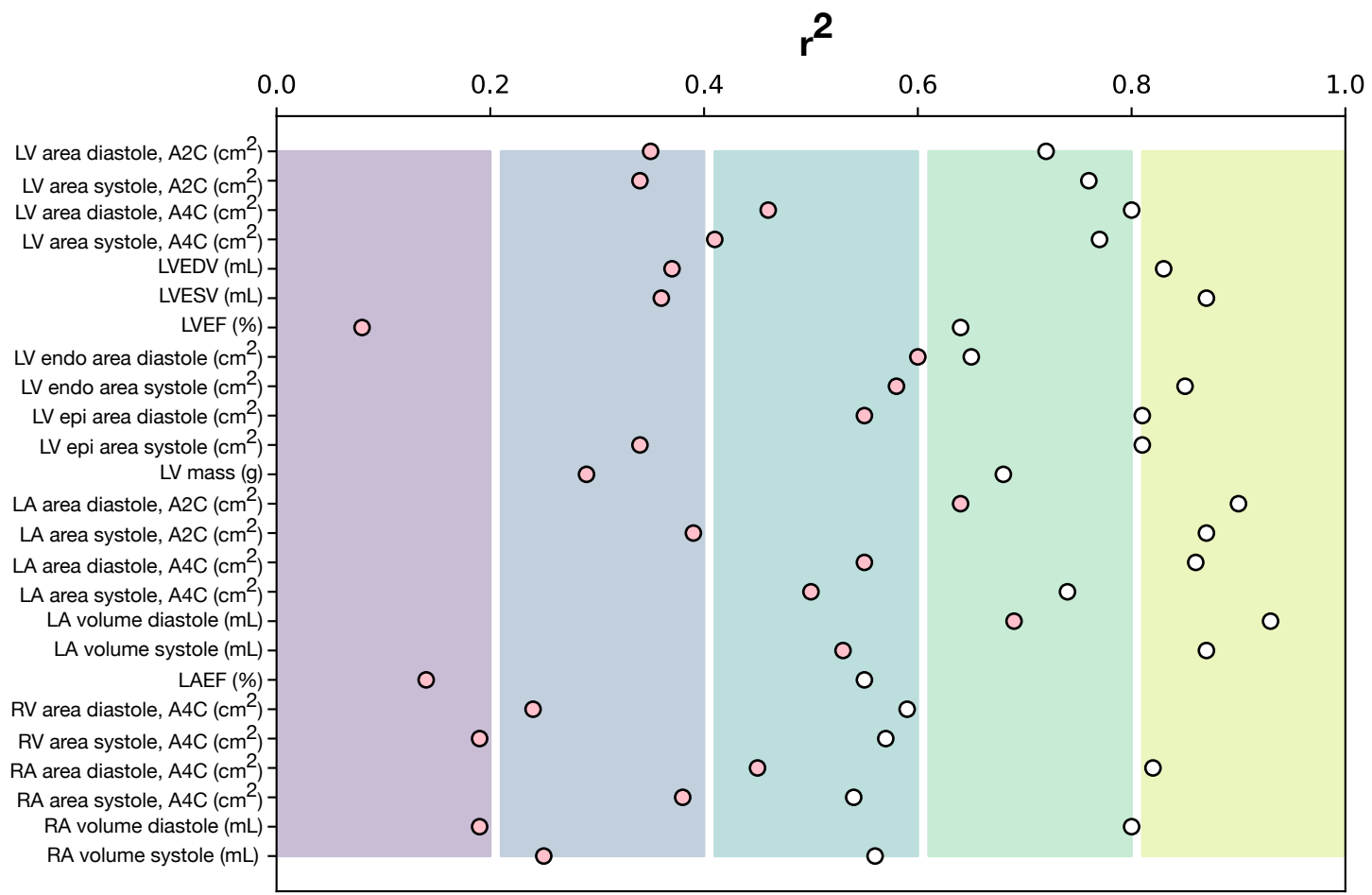

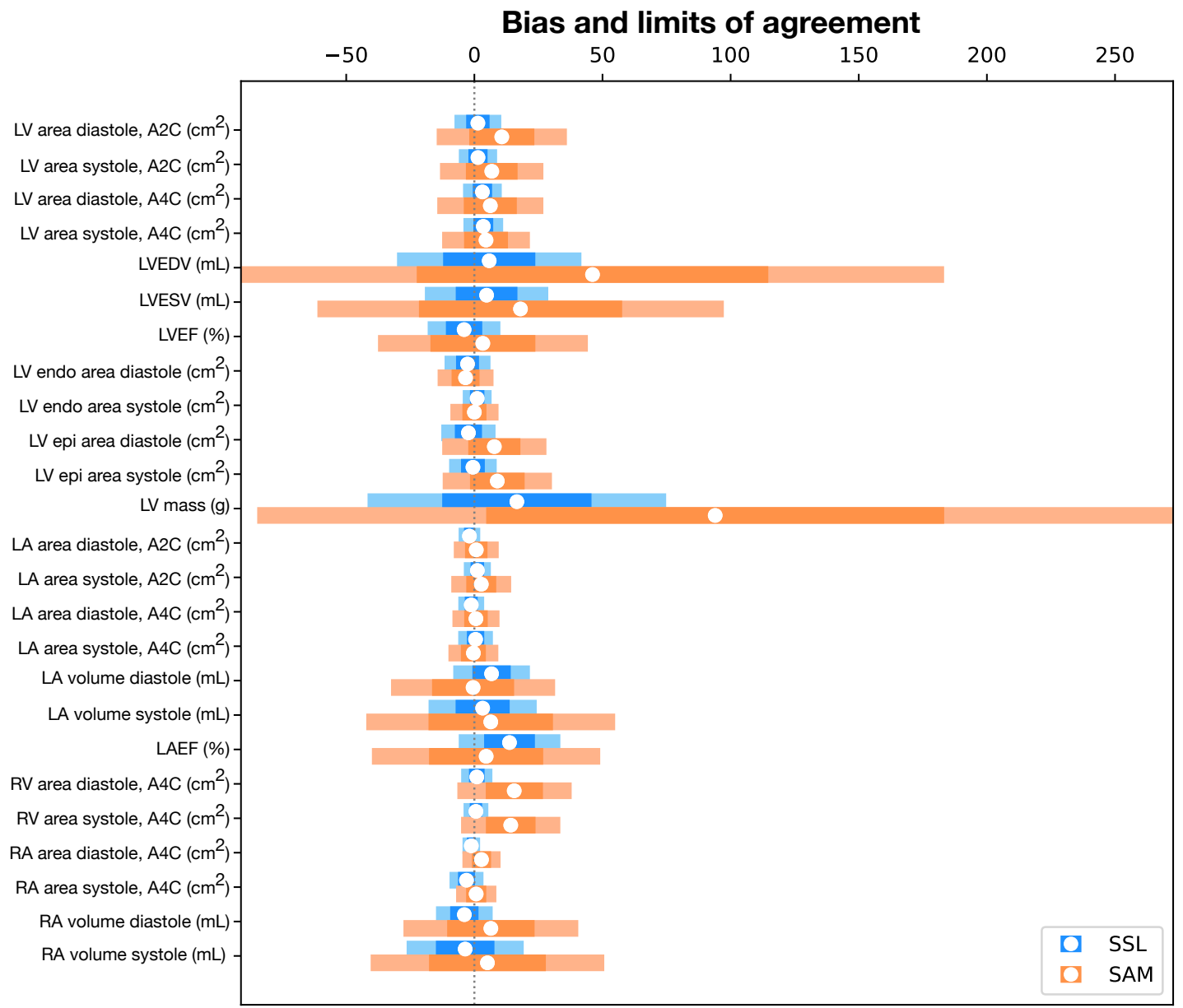

**Table S1. Demographics and clinical characteristics of the study population**

| Characteristics | n±std (range) or n(%) | Measurements | n±std (range) |
|---|---|---|---|
| Age, years | 61±17(24-89) | BSA, $m^2$ | 1.8±0.3 (1.3-2.7) |
| Female | 48 (48%) | LV Ejection Fraction, % | 59±12 (18-75) |
| LV Ejection Fraction <35% | 7 (7%) | LV End Diastolic Volume Index, $mL/m^2$ | 48±22 (21-135) |
| LV Ejection Fraction abnormal (qualitative) | 33 (33%) | LV End Systolic Volume Index, $mL/m^2$ | 21±18 (7-103) |
| LV Diastology abnormal | 46 (46%) | LV Mass Index, $g/m^2$ | 91±34 (34-290) |
| LV Size abnormal | 12 (12%) | LA Volume Index, $mL/m^2$ | 30±13 (14-91) |
| LV Mass abnormal | 28 (28%) | RA Volume, mL | 18±7 (6-40) |
| LA size abnormal | 33 (33%) | RV End Diastolic Area, $cm^2$ | 18±4 (8-29) |
| RA size abnormal | 17 (17%) | RV End Systolic Area, $cm^2$ | 9±4 (3-21) |
| RV size abnormal | 11 (11%) | | |
| RV function abnormal | 10 (10%) | | |

**Number of patients 100. LV = left ventricle, LA = left atrium, RV = right ventricle, RA= right atrium, BSA = body surface area.**